\documentclass[prd,twocolumn,floatfix,preprintnumbers,nofootinbib,nopacs]{revtex4}
\usepackage[utf8x]{inputenc}
\usepackage{mathrsfs}
\usepackage{amsmath,amsfonts,amssymb,amsthm, bm} %math stuff
\usepackage{graphicx}

\usepackage[caption = false]{subfig}
\usepackage[normalem]{ulem}
\usepackage{xcolor}
\definecolor{lcolor}{rgb}{0.5,0,0}
\definecolor{citcolor}{rgb}{0,0.3,0.0}
\usepackage{nicefrac}
\usepackage{comment}
\usepackage{tabularx}

\usepackage[breaklinks,colorlinks,urlcolor=blue,citecolor=citcolor,linkcolor=lcolor]{hyperref}
\usepackage{mciteplus}

\usepackage{physics}

\usepackage{enumitem}

\newcommand{\xt}{{\mathbf{x}}}

\newcommand{\yt}{{\mathbf{y}}}

\newcommand{\nc}{{N_\mathrm{c}}}

\newcommand{\cf}{C_\mathrm{F}}

\newcommand{\as}{\alpha_{\mathrm{s}}}
\newcommand{\aem}{\alpha_{\mathrm{em}}}

\makeatletter
\newcommand*{\rom}[1]{\expandafter\@slowromancap\romannumeral #1@}
\makeatother

\begin{document}
\author{Magnus Bertilsson}
\author{Tuomas Lappi}
\author{Heikki Mäntysaari}
\author{Xuan-Bo Tong}
\affiliation{
Department of Physics, University of Jyväskylä,  P.O. Box 35, 40014 University of Jyväskylä, Finland
}
\affiliation{
Helsinki Institute of Physics, P.O. Box 64, 00014 University of Helsinki, Finland
}

\title{The DIS dipole picture cross section at finite energy}

\begin{abstract}
We implement a finite energy constraint in the dipole picture of deep inelastic scattering, by restricting the invariant mass of the produced partonic system  by the virtual photon-target center of mass energy. We show that, for $Q^2=1$GeV$^2$, the effect of this constraint can reach up to $\sim$35\% for charm quarks and $\sim$7\% for light quarks at $x=0.01$, but then rapidly decreases at smaller $x$ or larger $Q^2$. 
\end{abstract}

\maketitle

\section{Introduction}
Deep inelastic scattering (DIS) experiments have been used since the late 1960's to study the structure of the proton by measuring its structure functions $F_L$ and $F_2$ by scattering electrons off protons~\cite{Bloom:1969kc}. The pointlike structure of the electron has enabled precision measurements  at facilities such as SLAC, HERA and JLab that have played a central role in uncovering the partonic structure of the proton.

At high collision energy, or  at small Bjorken $x$, the phenomenon of \textit{gluon saturation} has been predicted. At small $x$, the gluon density becomes sufficiently large that nonlinear effects of Quantum Chromodynamics (QCD) sets in,  taming the growth of the total cross section that would otherwise violate unitarity.  The dynamically generated scale at which saturation phenomena, e.g. gluon recombination, become important is referred to as the saturation scale $Q_s(x)$, and was first theorized in 1983 in Ref.~\cite{Gribov:1983ivg}. 

A convenient framework to describe saturation physics is provided by the
Color Glass Condensate (CGC)~\cite{McLerran:1993ka, McLerran:1993ni, Iancu:2003xm, Garcia-Montero:2025hys}, an effective theory of QCD at high energies. Although gluon saturation is a robust prediction of QCD, experimental signatures have so far remained elusive~\cite{Morreale:2021pnn}. With its high energy and luminosity, the upcoming Electron-Ion Collider~(EIC)~\cite{AbdulKhalek:2021gbh} is expected to provide unprecedented precision in probing the onset of gluon saturation phenomena. 

Proton structure functions play a central role when one tries to find definite signatures of gluon saturation. On one hand, precision measurements at HERA~\cite{H1:2013ktq, ZEUS:2014thn, H1:2015ubc, H1:2018flt} provide a stringent test of the gluon saturation dynamics described within the CGC. This will especially be the case in the next decade, when the Electron-Ion Collider will measure nuclear structure functions for the first time in collider kinematics. 
Nuclear targets are especially interesting because higher parton densities are expected to enhance saturation effects.
On the other hand, CGC calculations for  other processes, that provide complementary probes to the saturated part of the proton and nuclear wave functions, require some non-perturbative input. This is typically obtained by inferring an initial condition for  high-energy evolution equations from  proton structure function data~\cite{H1:2013ktq, ZEUS:2014thn, H1:2015ubc, H1:2018flt}. Many dipole picture fits to HERA total cross section data have been performed~\cite{Albacete:2009fh, Albacete:2010sy, Casuga:2023dcf, Gao:2025dkn, Mantysaari:2018zdd, Beuf:2020dxl,Hanninen:2025iuv, Hanninen:2022gje,Casuga:2025etc}. However, the inclusive cross section data alone has turned out to be inconclusive in clearly establishing an experimental gluon saturation, see e.g. Ref.~\cite{Mantysaari:2018nng}. 

At small $x$, the DIS cross section is most commonly formulated within the CGC framework which leads to the dipole picture~\cite{Nikolaev:1990ja, Nikolaev:1991et,Mueller:1994jq} of DIS. In this approach, the scattering amplitude can be factorized into two terms: the splitting of the virtual photon $\gamma^*$ into a quark-antiquark dipole (at lowest order), and the subsequent interaction of this dipole with the target as described by CGC.
Within this framework, perturbative calculations have been performed up to next-to-leading order in $\as$ in~\cite{Beuf:2017bpd,Hanninen:2017ddy,Beuf:2021srj,Beuf:2022ndu,Beuf:2021qqa} (see also Refs.~\cite{Beuf:2020dxl, Hanninen:2022gje,Casuga:2025etc} for phenomenological applications). These higher-order corrections are crucial in order to match the precision achieved at HERA and at the EIC. 

The total cross section is typically obtained by invoking the optical theorem, which allows one to express the total cross section in terms of the forward elastic dipole-target scattering amplitude.  
This implicitly assumes an asymptotically large center-of-mass energy, $W$, and that the phase space of the produced partonic system  at the unitarity cut is unconstrained. Away from this strict high-energy limit, this assumption is no longer exact, and corrections to the usual dipole picture result arise from the fact that the invariant mass of the produced partonic system ($M_{q\Bar{q}}$ for the $q\bar q$ system at leading order), can at most be equal to $W$. While various kinematical constraints have been discussed in the literature related to the small-$x$ evolution~\cite{Kuokkanen:2011je,Beuf:2014uia, Watanabe:2015tja} and as corrections to the eikonal interaction with the target~\cite{Chirilli:2018kkw,Chirilli:2021lif,Agostini:2024xqs,Altinoluk:2024zom,Altinoluk:2025ivn,Agostini:2025vvx}, the impact of imposing a finite-energy constraint on the invariant mass within the dipole-picture cross section has not been systematically explored. 

We show that it is possible to implement a finite energy correction to the DIS process within the dipole picture while maintaining the essential dipole picture in the sense that scattering is given by lightlike Wilson lines, and at LO in $\as$ the DIS process consists of the scattering of a $q\Bar{q}$-dipole off the target. This is done without introducing subeikonal interactions and is fully consistent with the way the shockwave scattering picture (often also referred to as dipole picture) is used for differential observables (forward hadron production~\cite{Chirilli:2011km}) and exclusive and diffractive DIS~\cite{Mantysaari:2021ryb,Beuf:2024msh}, where one measures the final state.Additionally, we propose that including this invariant mass constraint, which removes the nonphysical part of phase space, will improve the precision of the DIS dipole picture calculations. In fact what we are doing is bringing the description of the DIS process in line to be consistent with all these other applications of the same formalism to different scattering processes.

The paper is organized as follows. In Sec.~\ref{sec:Theory} we calculate the total DIS cross section in the dipole picture including a finite energy (FE) constraint,  keeping the maximum invariant mass arbitrary at this point. This is followed by a numerical analysis: the model used to describe the interaction with the target shockwave is described in 
Sec.~\ref{sec:Quadrupole} and numerical results quantifying the impact of the finite energy constraint, with the maximum invariant mass now fixed to the center-of-mass energy of the photon-target collision system, are shown in Sec.~\ref{sec:Results}. Finally in Sec.~\ref{sec:Conclusion}  we draw our conclusions and present an outlook of future work.

\section{Inclusive DIS at finite energy}\label{sec:Theory}
\begin{figure}[t]
    \centering
    \includegraphics[width=\linewidth]{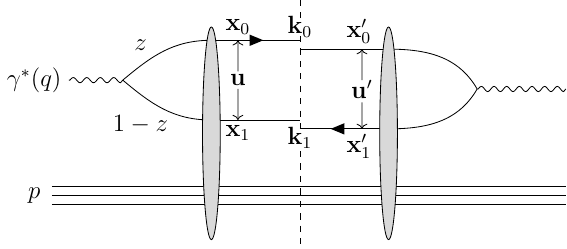}
    \caption{Diagram representing the $\gamma^* p$ cross-section. The left and right hand sides represent the amplitude and its conjugate, respectively. 
    }
    \label{fig:QuadrupoleDiagram}
\end{figure}

Our aims in this paper are to
\begin{enumerate}[label=\alph*)]
    \item derive the leading order (LO) DIS cross section in the dipole picture including explicit limits for the final state kinematics, and
    \item numerically quantify the importance of this finite energy constraint.
\end{enumerate} 
Under the one-photon approximation, the inclusive DIS structure functions can be written in terms of the virtual photon-proton cross sections $\sigma^{\gamma^* p}_{L,T}$ as 
\begin{align}
    &F_L(x, Q^2) = \frac{Q^2}{4\pi^2 \aem} \sigma^{\gamma^* p}_L \\
    &F_2(x, Q^2) = \frac{Q^2}{4\pi^2 \aem} \left(\sigma^{\gamma^* p}_L + \sigma^{\gamma^* p}_T \right).
\end{align}
Here \textit{L} and \textit{T} respectively refer to the  \textit{longitudinal} and \textit{transverse} polarizations of the $\gamma^*$, and $\aem$ is the fine structure constant. The photon virtuality is given by $Q^2 = -q^2$ which is related to Bjorken  $x = Q^2 / (2p\cdot q)$, where $p$ and $q$ respectively are the proton and virtual photon four-momenta. In the leading order dipole picture, the system produced from the photon in the final state is a $q\bar{q}$ dipole. The cross section differentially in the quark and antiquark momenta is conventionally referred to as the ``dijet'' cross section, even if the quarks do not have a particularly high transverse momentum.
Integrating this dijet production cross section in DIS over the final state kinematics, gives the total DIS cross section as~\cite{Dominguez:2011wm}
\begin{equation}\label{eq:generic_cs_dip_coords}
    \begin{split}
        \sigma^{\gamma^* p}_{L,T} &= \int \frac{\dd[2]\mathbf{K}}{(2\pi)^2}\, \frac{\dd[2]\mathbf{P}}{(2\pi)^2}\dd{z} \dd[2]{\mathbf{u}} \dd[2]{\mathbf{v}} \dd[2]{\mathbf{u}^\prime} \dd[2]{\mathbf{v}^\prime} \\
        & \cross e^{i\mathbf{K} \cdot (\mathbf{v}^{\prime} - \mathbf{v})}e^{i\mathbf{P} \cdot (\mathbf{u}^{\prime} - \mathbf{u})} \, \mathcal{N}(z,\,\mathbf{u},\,\mathbf{v},\,\mathbf{u}^{\prime},\,\mathbf{v}^{\prime})\\
        &\cross \Psi_{L,T}^{\gamma^* q\Bar{q}}(z,\,\abs{\mathbf{u}})\left(\Psi_{L,T}^{\gamma^* q\Bar{q}}(z,\,\abs{\mathbf{u}^\prime})\right)^\dagger.
    \end{split}
\end{equation}
Schematically, this expression can be depicted as in Fig.\,\ref{fig:QuadrupoleDiagram}. Here $\Psi_{L,T}^{\gamma^* q\Bar{q}}$ and $(\Psi_{L,T}^{\gamma^* q\Bar{q}})^\dagger$ are the photon lightcone wavefunctions (LCWFs) in the amplitude and conjugate amplitude, respectively, which describe the $\gamma^*$ splitting into a $q\Bar{q}$-pair. The quark and antiquark respectively carry a fraction $z$ and $(1-z)$ of the large photon plus momentum, $q^+$. The transverse size of the dipole in the amplitude is $\mathbf{u}$ and its center-of-mass $\mathbf{v}$. Primed vectors such as $\mathbf{u}^\prime$ and $\mathbf{v}^\prime$ refer to the corresponding coordinates in the conjugate amplitude. 
These vectors are related to the quark (antiquark) transverse positions $\xt_0$ ($\xt_1$) as 
\begin{equation}
    \begin{split}
        \mathbf{x}_0 &= (1-z)\mathbf{u} + \mathbf{v}, \quad\quad \mathbf{x}_0^\prime = (1-z)\mathbf{u}^\prime + \mathbf{v}^\prime \\
        \mathbf{x}_1 &= -z\mathbf{u} + \mathbf{v}, \quad \quad \quad ~~\mathbf{x}_1^\prime = -z\mathbf{u}^\prime + \mathbf{v}^\prime.
    \end{split}
\end{equation}

The final state momentum variables  are the total transverse momentum of the dipole $\mathbf{K}$ (i.e. total momentum transfer from the target) and the momentum imbalance $\mathbf{P}$. They are related to the final state quark and antiquark transverse momenta $\mathbf{k}_0$ and $\mathbf{k}_1$ as
\begin{equation}
    \begin{split}
        \mathbf{P} &= (1-z)\mathbf{k}_0 - z\mathbf{k}_1, \quad\quad \mathbf{K} = \mathbf{k}_0 + \mathbf{k}_1 .
    \end{split}
\end{equation}
In the cross section~\eqref{eq:generic_cs_dip_coords}, factors of $\frac{1}{4\pi}$ and $\frac{1}{z (1-z)}$ are commonly seen. We absorb these into the light cone wavefunctions, which are now given by 
\begin{align}
    &\Psi_{L}^{\gamma^* q\Bar{q}}(z,\,\abs{\mathbf{u}})\left(\Psi_{L}^{\gamma^* q\Bar{q}}(z,\,\abs{\mathbf{u}^\prime})\right)^\dagger = \nc \sum_f \frac{
    \aem Q_f^2}{4\pi^2}\\ & \times 8Q^2 z^2(1-z)^2 K_0(\varepsilon_f \abs{\mathbf{u}})K_0^*(\varepsilon_f \abs{\mathbf{u}^\prime}) \nonumber \\
    & \Psi_{T}^{\gamma^* q\Bar{q}}(z,\,\abs{\mathbf{u}})\left(\Psi_{T}^{\gamma^* q\Bar{q}}(z,\,\abs{\mathbf{u}^\prime})\right)^\dagger = 2\nc \sum_f \frac{\aem Q_f^2}{4\pi^2} \\ &\times \big[(z^2 + (1-z)^2)\varepsilon_f^2 \cos{(\alpha)}K_1(\varepsilon_f \abs{\mathbf{u}})K_1^*(\varepsilon_f \abs{\mathbf{u}^\prime})\nonumber\\
    & + m_f^2 K_0(\varepsilon_f \abs{\mathbf{u}})K_0^*(\varepsilon_f\abs{\mathbf{u}^\prime})\big]. \nonumber
\end{align}
The wave function overlaps have already been summed over quark helicities and, in the case of the transverse photon, averaged over the polarization states.
Here $f$ labels the quark flavor, $\nc$ is the number of colors, $m_f$ is the quark mass, $\varepsilon_f^2 = Q^2z(1-z) + m_f^2$, $Q_f$ is the quark fractional charge, $\alpha$ is the angle between $\mathbf{u}$ and $\mathbf{u}^\prime$, and $K_0$ and $K_1$ are modified Bessel functions of the second kind. 
For brevity, the magnitude of a 2D vector will from now on be denoted by $\abs{\mathbf{a}} = a_\perp$.

The function $\mathcal{N}$ in Eq.~\eqref{eq:generic_cs_dip_coords} describes the eikonal interaction of the quarks with the target. It is given by 
\begin{equation}\label{eq:TargetAmp}
    \mathcal{N}(z,\,\mathbf{u},\,\mathbf{v},\,\mathbf{u}^{\prime},\,\mathbf{v}^{\prime}) = S^{(4)} - S^{(2)} - S^{\prime(2)} + 1, 
\end{equation}
where
\begin{equation}\label{eq:QpoleandDipoles}
    \begin{split}
        & S^{(4)} = \frac{1}{\nc}\left\langle \Tr{U(\mathbf{x}_0)U^\dagger(\mathbf{x}_1)U(\mathbf{x}_1^\prime)U^\dagger(\mathbf{x}_0^\prime)}\right\rangle \\
        & S^{(2)} = \frac{1}{\nc} \left\langle \Tr{U(\mathbf{x}_0)U^\dagger(\mathbf{x}_1)}\right\rangle \\
        & S^{\prime(2)} = \frac{1}{\nc} \left\langle \Tr{U(\mathbf{x}_1^\prime)U^\dagger(\mathbf{x}_0^\prime)}\right\rangle,
    \end{split}
\end{equation}
are correlators of Wilson lines, $U(\mathbf{x})$, dependent on the transverse positions of the individual quarks. The Wilson lines are expressed in the fundamental representation \cite{Kovchegov:2012mbw}
\begin{equation}
    U(\mathbf{x}) = \mathcal{P}\text{exp}\left(-ig \int \dd{x^+} A^-_a(x^+, \mathbf{x})t^a\right)
\end{equation}

Previous calculations of the proton structure functions in the dipole picture have  used the \textit{optical theorem} (OT), which is equivalent to integrating the dijet cross section over the final state kinematics ($\mathbf{P},\mathbf{K},z$) without any finite energy constraint.
Integrating Eq.~\eqref{eq:generic_cs_dip_coords}
over  $\mathbf{K}$  sets the impact parameters to be the same in the amplitude and conjugate amplitude, $\mathbf{v}=\mathbf{v}^\prime$. Assuming a translationally invariant target, the Wilson line correlators become independent of $\mathbf{v}$. Under this assumption, integration over $\mathbf{v}$ only contributes an overall factor $\sigma_0/2$ corresponding to the transverse size of the target:
\begin{equation}\label{eq:translationInvarinat}
    \int \dd[2]{\mathbf{v}} \xrightarrow{} \frac{\sigma_0}{2}.
\end{equation}

\begin{figure}[bt]
    \centering
    \includegraphics[width=0.8\linewidth]{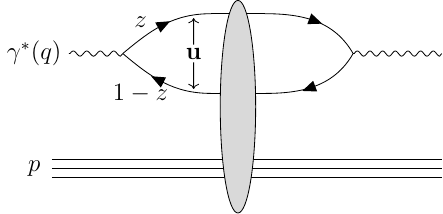}
    \caption{Diagram representing the forward elastic scattering amplitude for $\gamma+p\to\gamma+p$, used to compute the total cross section obtained through the optical theorem.}
    \label{fig:OTdiagram}
\end{figure}

Then, the integration over  $\mathbf{P}$ sets the dipole sizes to be identical in the amplitude and conjugate amplitude and reduces the quadrupole $S^{(4)}$ to $1$.
This results in the well known expression typically derived using the optical theorem:
\begin{equation}\label{eq:OT_cs}
    \begin{split}
        \sigma^{\gamma^* p}_{L,T} &= \sigma_0\int \dd{z} \dd[2]{\mathbf{u}}\, N(\mathbf{u})\left|\Psi_{L,T}^{\gamma^* q\Bar{q}}(z,\,u_\perp)\right|^2
    \end{split}
\end{equation}
depicted in Fig.~\ref{fig:OTdiagram}. Here 
\begin{equation}\label{eq:DipoleAmp}
    N(\mathbf{u}) = 1 - S^{(2)}.  
\end{equation}
We will denote the structure functions calculated using this result as $F_L^{OT}$ and $F_2^{OT}$. They will serve as a benchmark to compare our finite energy corrected cross section. 

We implement the finite scattering energy correction by requiring that the invariant mass of the produced $q\Bar{q}$-system $M_{q\Bar q}$ must be less than or equal to the center-of-mass energy, $W$, of the $\gamma^* p$-system, i.e.
\begin{equation}\label{eq:Kin_constraint}
    M^2_{q\Bar{q}} \leq W^2 = (q + p)^2 \approx \frac{Q^2}{x},
\end{equation} 
where the last approximation is valid at small $x$. The invariant mass for a quark-antiquark pair of flavor $f$ and mass $m_f$ is given by 
\begin{equation}
    \begin{split}
        M_{q\Bar{q}}^2 &= (k_0 + k_1)^2 = 2m_f^2 + 2(k_0 \cdot k_1) \\
        & = 2m_f^2 + k_0^+k_1^- + k_0^-k_1^+ - 2\mathbf{k}_0\cdot\mathbf{k}_1
    \end{split}
\end{equation}
In terms of the longitudinal momentum fraction $z$, the invariant mass becomes
\begin{equation}\label{eq:invariant_mass_TRF}
    \begin{split}
        M_{q\Bar{q}}^2 & = m_f^2 + \frac{\mathbf{k}_0^2}{z} + \frac{\mathbf{k}_1^2}{(1-z)} - (\mathbf{k}_0 + \mathbf{k}_1)^2 \\
        &  = \frac{\mathbf{P}^{2} + m_f^2}{z(1-z)}.
    \end{split}
\end{equation}
Here we see that limiting the invariant mass $M_{q\bar q}^2$ by the center-of-mass energy $W$, effectively limits the phase space of the $\mathbf{P}$-integral. It does not limit the phase space of $\mathbf{K}$, as the overall transverse momentum of the quark pair does not contribute to its invariant mass. 
Consequently, integration over the momentum transfer again sets $\mathbf{v}=\mathbf{v'}$.

The integral over $\mathbf{P}$ in Eq.~\eqref{eq:generic_cs_dip_coords} is of the form
\begin{equation}
    \begin{split}
        I_P(\mathbf{u},\,\mathbf{u}^\prime) = &\int \frac{\dd[2]{\mathbf{P}}}{(2\pi)^2}e^{i\mathbf{P} \cdot (\mathbf{u}^{\prime} - \mathbf{u})}\\
        = \frac{1}{2\pi}&\int_0^{P^{\mathrm{max}}_\perp} \dd{P_\perp} P_\perp J_0(P_\perp\abs{\mathbf{u}^{\prime} - \mathbf{u}}).
    \end{split}
    \label{eq:Ipdef}
\end{equation}
We implement the constraint by introducing
\begin{equation}\label{eq:kinematic_constraint}
    \begin{split}
        1 = \int_{M^2_{\mathrm{min}}}^{M_{\mathrm{max}}^2} \dd{M}^2_{q\Bar{q}}\, \delta\left(M^2_{q\Bar{q}} - \frac{P_{\perp}^2 + m_f^2}{z(1-z)}\right)
    \end{split}
\end{equation}
into $I_P$, where $P_{\perp} < P_{\perp}^{\mathrm{max}}$ and $P_{\perp}^{\mathrm{max}}$ is set by the final state kinematics. It turns out to be simpler to perform numerical calculations if one changes variables to $M^2_{q\Bar{q}}$ by integrating over $\mathrm{P}_\perp$. The integral bounds are 
\begin{align}
    M_{\mathrm{max}}^2 & = \frac{(P^{\mathrm{max}}_\perp)^2 + m_f^2}{z(1-z)} \\
    M_{\mathrm{min}}^2 & = \frac{m_f^2}{z(1-z)}.
\end{align}
These bounds follows directly from the change of variables. Note that the optical theorem result~\eqref{eq:OT_cs} can be obtained in the limit $M_\mathrm{max}^2\to \infty$. Thus the finite energy constraint in Eq.~(\ref{eq:Kin_constraint}) will have an increasing effect at lower values of $M_{\mathrm{max}}^2$. Since this upper limit  should be taken as $W^2$, the effect of the constraint will grow towards increasing $x$ for a fixed $Q^2$ or decreasing $Q^2$ at fixed $x$. 

With this change of variables, Eq.~\eqref{eq:Ipdef} can be written as
\begin{equation}
\begin{split}\label{eq:invMassintegral}
    \widetilde{I}_P(M_{\mathrm{max}}^2;&\,z,\,\mathbf{u},\,\mathbf{u}^\prime)  = \frac{1}{4\pi} \int_{M^2_{\mathrm{min}}}^{M_{\mathrm{max}}^2} \dd{M}_{q\Bar{q}}^2\, z(1-z)\\
    & \cross J_0\left(\abs{\mathbf{u}^{\prime} - \mathbf{u}}\sqrt{M^2_{q\Bar{q}}z(1-z) - m_f^2}\right).
\end{split}
\end{equation}
One can perform the \(M_{q\Bar{q}}^2\)-integral analytically to get 
\begin{equation}
    \begin{split}
       \widetilde{I}_P(M^2_{\mathrm{max}};\,z,\,\mathbf{u},\,\mathbf{u}^\prime)& = \frac{\zeta J_1 \left(\zeta\right)}{2\pi\abs{\mathbf{u}^{\prime} - \mathbf{u}}^2}.
    \end{split}
\end{equation}
%where ${}_0\Tilde{F}_1$ is a regularized hypergeometric function\footnote{In our case ${}_0\Tilde{F}_1(2,\zeta)$ = ${}_0{F}_1(2,\zeta)$, the \textit{confluent hypergeometric function} which is available in the  \href{https://docs.scipy.org/doc/scipy/reference/generated/scipy.special.hyp2f1.html}{SciPy library}~\cite{2020SciPy-NMeth}.}\,\cite{article:HyperGeom} 
The argument of the Bessel functoin $J_1$ is given by 
\begin{align}
   \zeta = \sqrt{M_\mathrm{max}^2z(1-z) - m_f^2}\abs{\mathbf{u}^{\prime} - \mathbf{u}}.
\end{align}
As $M^2_{\mathrm{max}} \to \infty$,  $J_1(\zeta)$ becomes an increasingly rapidly oscillating function, so that all contributions except $\mathbf{u}^{\prime} \approx \mathbf{u}$ cancel in the integral. This oscillatory nature requires care in the numerical integration procedure.

Compared to the optical theorem result~\eqref{eq:OT_cs}, where one has integrated over $\mathbf{P}$ with no constraint setting $\mathbf{u}=\mathbf{u}^\prime$, now the trace over the Wilson lines in the quadrupole $S^{(4)}$ given in Eq.~\eqref{eq:QpoleandDipoles} does not reduce to unity. We again consider a large uniform target\footnote{The importance of finite-size effects has been recently highlighted in Ref.~\cite{Mantysaari:2024zxq}}, and consequently the Wilson line correlators  are independent of $\mathbf{v}$. In this case these correlators can be written as
\begin{equation}\label{eq:QpoleandDipoles_ucoords}
    \begin{split}
        & S^{(4)} \propto \left\langle \Tr{U((1-z)\mathbf{u})U^\dagger(-z\mathbf{u})U(-z\mathbf{u}^\prime)U^\dagger((1-z)\mathbf{u}^\prime)}\right\rangle \\
        & S^{(2)} \propto \left\langle \Tr{U((1-z)\mathbf{u})U^\dagger(-z\mathbf{u})}\right\rangle \\
        & S^{\prime(2)} \propto \left\langle \Tr{U(-z\mathbf{u}^\prime)U^\dagger((1-z)\mathbf{u}^\prime)}\right\rangle,
    \end{split}
\end{equation}
and the integration over $\mathbf{v}$ gives an overall factor of the proton transverse area, see Eq.~\eqref{eq:translationInvarinat}.
With the notation
\[\frac{\sigma_0}{2} \widetilde{\mathcal{N}}(z,\,\mathbf{u},\,\mathbf{u}^{\prime}) = \int \dd[2]{\mathbf{v}} \mathcal{N}(z,\,\mathbf{u},\,\mathbf{u}^{\prime},\,\mathbf{v})\]
we can now write the cross section as 
%The only remaining angular dependence in Eq.\,(\ref{eq:generic_cs_dip_coords}) is on the relative angle, $\alpha$, between the two vectors \(\mathbf{u}\) and $\mathbf{u}^{\prime}$. Therefore, the cross section can be integrated over one overall angle and written as
\begin{equation}\label{eq:Final_cs}
    \begin{split}
        \sigma^{\gamma^* p}_{L,T} &= \frac{\sigma_0}{2} \int \dd{z}  \dd[2]{\mathbf{u}}\dd[2]{\mathbf{u}^\prime}\Psi_{L,T}^{\gamma^* q\Bar{q}}(z,\,u_\perp) \left(\Psi_{L,T}^{\gamma^* q\Bar{q}}(z,\,u_\perp^\prime)\right)^\dagger \\
        &\times \widetilde{\mathcal{N}}(z,\,\mathbf{u},\,\mathbf{u}^{\prime})\widetilde{I}_P(M^2_{\mathrm{max}};\,z,\mathbf{u},\,\mathbf{u}^{\prime}). 
    \end{split}
\end{equation}
In contrast to the structure functions obtained through the optical theorem~\eqref{eq:OT_cs}, we will denote the structure functions calculated using the finite energy constraint by $F_L^{FE}$ and $F_2^{FE}$. The DIS cross section with the finite energy correction given in Eq.~\eqref{eq:Final_cs} is our main result a) stated at the beginning of this Section.

\section{Modeling the correlators}\label{sec:Quadrupole}
In Eq.~\eqref{eq:Final_cs} it is no longer possible to perform the remaining integrals analytically. To this end we have developed a numerical implementation \cite{Bertilsson_Precision_probes_EIC_2025} which uses an adaptive Vegas Monte Carlo algorithm to numerically integrate over the  phase space. Evaluating the integrand requires us to specify a model for the Wilson line correlators appearing in Eq.~\eqref{eq:QpoleandDipoles_ucoords}. 

As discussed above, the finite-energy corrected cross section involves a more complicated four-point Wilson line correlator $S^{(4)}$ (quadrupole), in addition to the common dipole $S^{(2)}$ correlator that is the only structure appearing in the optical theorem result~\eqref{eq:OT_cs}. 
We evaluate the quadrupole by using the Gaussian approximation\footnote{The importance of having a realistic description of the quadrupole has been highlighted e.g. in Ref.~\cite{Lappi:2012nh} in the case of forward dihadron production in proton-nucleus collisions.} following Ref.~\cite{Dominguez:2011wm}, which allows one to express the quadrupole in terms of the dipole. For completeness we also calculated the cross section in the large-$\nc$ limit of the Gaussian approximation for the quadrupole, given also in \cite{Dominguez:2011wm}. The results obtained are identical within our numerical accuracy. Explicit expressions for the Gaussian approximation of the quadrupole can be found from Ref.~\cite{Dominguez:2011wm} and are also given in our notation in Appendix~\ref{sec:Appendix}. 

The Gaussian approximation is exact if the color charge distribution is Gaussian, as is the case in the McLerran-Venugopalan (MV) model~\cite{McLerran:1993ka, McLerran:1993ni}.
The MV model dipole used in this work reads
\begin{equation}
\label{eq:dipole}
    S^{(2)}(r) = e^{-\frac{1}{4}\left(r^2 Q_{s0}^2 \right)^\gamma \log\left( 1/(r\Lambda_\mathrm{QCD}) + e\cdot e_c\right)}.
\end{equation}
In principle the dipole amplitude (and higher point correlators) satisfy the Balitsky-Kovchegov (BK)~\cite{Kovchegov:1999yj,Balitsky:1995ub} or JIMWLK evolution equations~\cite{Jalilian-Marian:1997qno,Jalilian-Marian:1997ubg,Kovner:2000pt,Iancu:2000hn,Mueller:2001uk}, which determine their dependence on  Bjorken $x$. 
Since the purpose of this work is to estimate the importance of the finite energy constraint rather than to compare with, for example, HERA data, we neglect the BK/JIMWLK evolution, and consider the Wilson line correlators to be independent of $x$. Also, the correction is largest at large $x$ where the correlators are still close to the initial condition.
We expect the relative importance of the finite energy correction to depend only weakly on the actual form of the dipole. In general, BK evolution drives the dipole towards an asymptotic solution with smaller anomalous dimension ($\gamma<1$), corresponding to a harder spectrum in momentum space and larger invariant masses. Therefore, the results presented in this work can be regarded as lower bounds on the estimated importance of the finite energy correction.

\begin{table*}[tb!]
    \centering
    \includegraphics[width=0.6\linewidth]{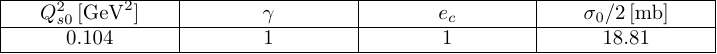}
    \caption{Parameters for the MV model dipole~\eqref{eq:DipoleAmp} obtained in Ref.~\cite{Lappi:2013zma} and used in numerical analysis.}
    \label{tab:params}
\end{table*}

The accuracy of the Gaussian approximation at small $x$ after the JIMWLK evolution has been demonstrated in Ref.~\cite{Dumitru:2011vk} for two specific coordinate space configurations. Although there could be larger effects for other coordinate configurations and Wilson line correlators, we still expect the Gaussian approximation to provide a good enough estimate for the operators needed to evaluate Eq.~\eqref{eq:Final_cs}. In particular, it is important that the Gaussian approximation preserves the ``coincidence limits''~\cite{Kovchegov:2008mk} following from its  Wilson line operator definition, where the quadrupole reduces to a dipole when one quark and one antiquark coordinate are the same.

For the numerical values presented in this paper, we use the initial condition from the BK evolution fit with the MV initial condition from Ref.~\cite{Lappi:2013zma}, where one fixes $\gamma=1$ and $e_c=1$. The parameters are given in Table~\ref{tab:params}. Since the fit uses the optical theorem cross section, the calculation with our correction will not precisely match HERA data even at $x=0.01$. However, we are here more interested in quantifying the relative importance of the finite energy correction. In the future it would be interesting to include our correction in BK fits to HERA data.

\begin{figure*}[tb!]   % or [htbp]
    \centering
    \subfloat{\includegraphics[width=0.5\textwidth]{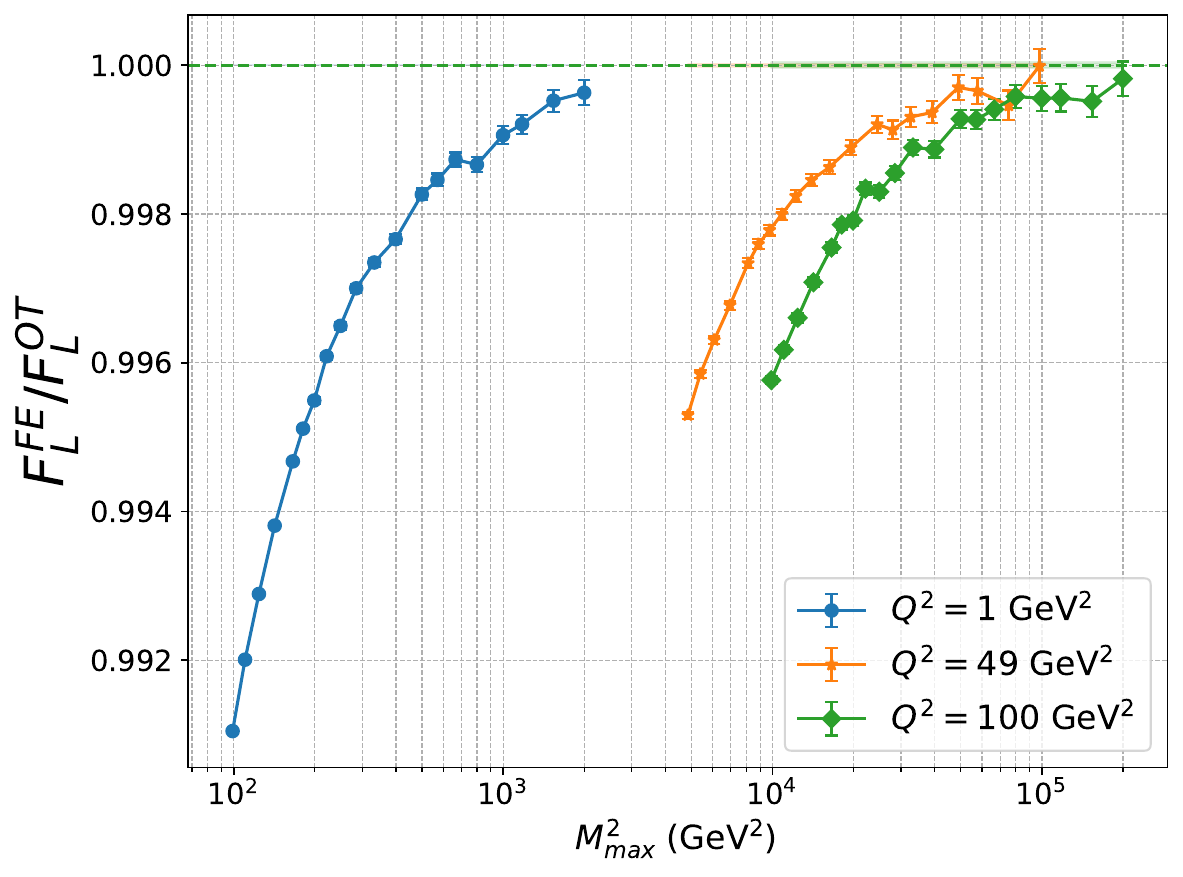}}\hfill
    \subfloat{\includegraphics[width=0.5\textwidth]{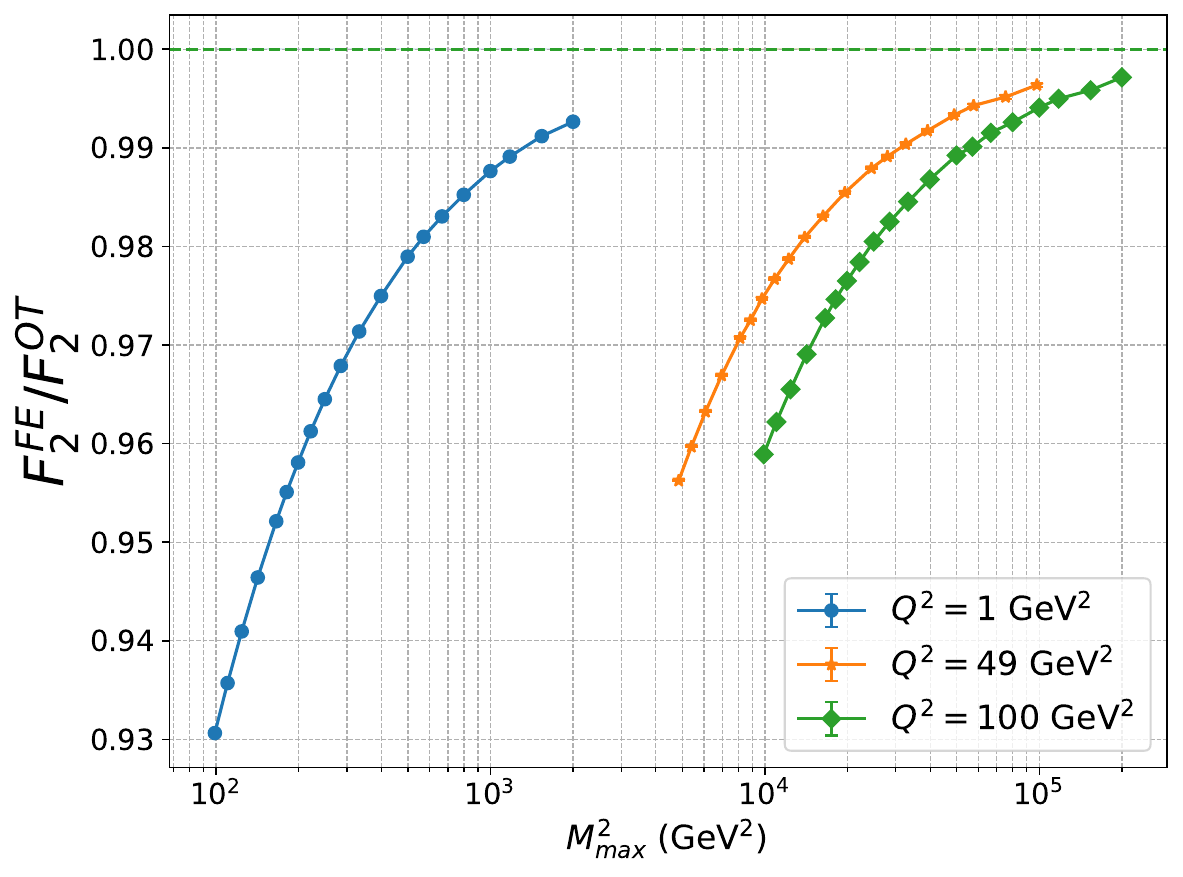}}
    \caption{Proton structure function $F_L$ (left) and $F_2$ (right) calculated with the finite energy correction, normalized by the optical theorem result.
  The uncertainty bars indicate the estimated numerical integration uncertainty.
    }
    \label{fig:Light1Dplot}
\end{figure*}
\begin{figure*}[tb!]   % or [htbp]
    \centering
    \subfloat{\includegraphics[width=0.5\textwidth]{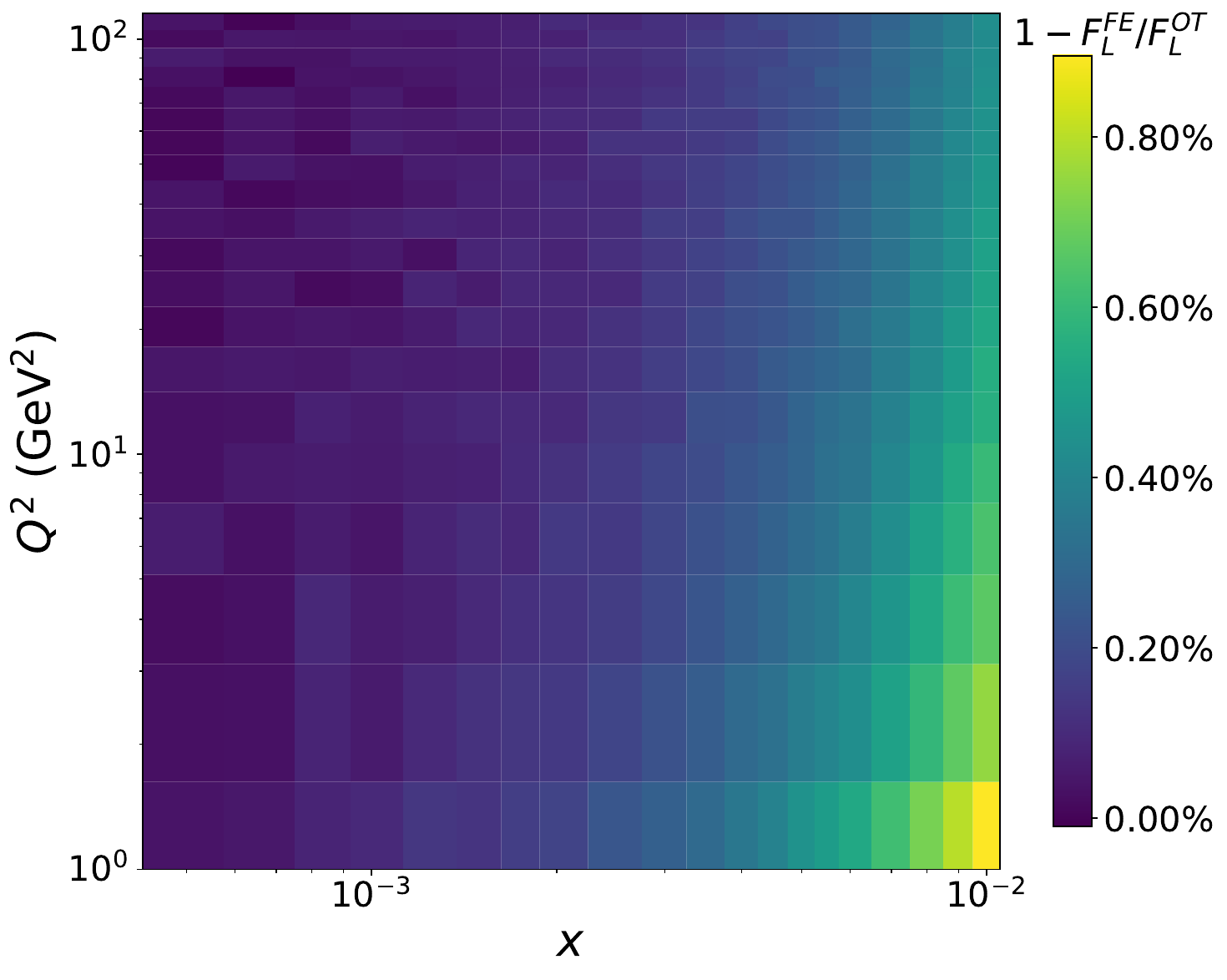}}\hfill
    \subfloat{\includegraphics[width=0.5\textwidth]{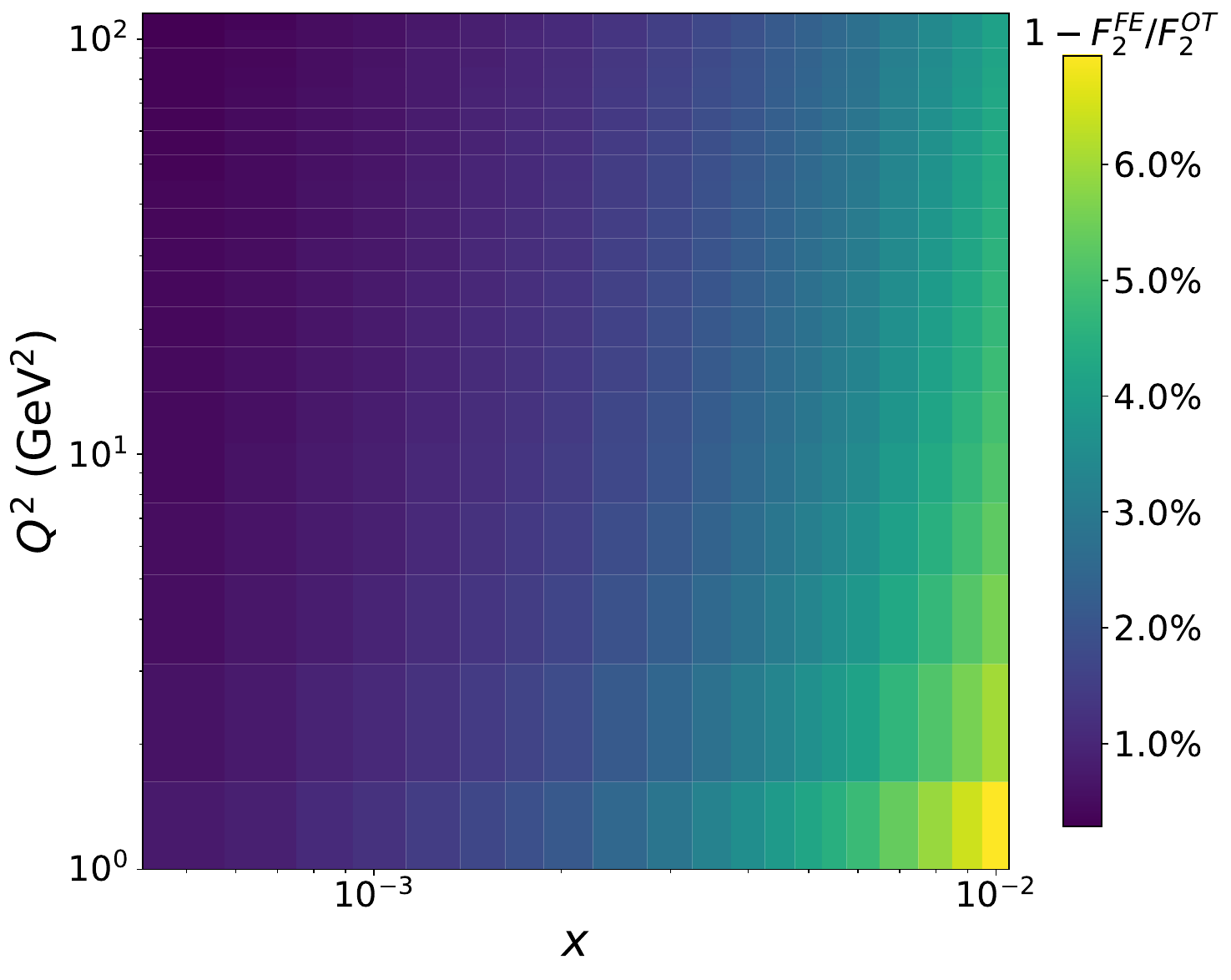}}
    \caption{Phenomenological impact of the finite energy correction on $F_L$ (left) and $F_2$ (right) in the $x$-$Q^2$ plane for light quarks.}
    \label{fig:LightHeatmap}
\end{figure*}

\section{Results}\label{sec:Results}
We determine the numerical effect of the finite energy correction on   proton structure functions. These results depend on the quark mass, and therefore we consider light and heavy (charm) quark production separately. In each case we first illustrate how the structure functions behave as a function of the upper limit of the $M^2_{q\Bar{q}}$-integral. We then quantify the impact of the finite energy correction over the $x\text{-}Q^2$ region covered at HERA and at the EIC: $1 < Q^2\lesssim 100\,\mathrm{GeV}^2$ and $10^{-4}\lesssim x < 0.01$. The lower limit in the $Q^2$ range justifies perturbative calculations and the upper limit for the $x$ range is a value that is typically used in phenomenological applications to guarantee applicability of the dipole picture.

\subsection{Light quarks} 
For the light quarks, we use a quark mass $m_f = 0.14$ GeV, a value that is often used in fits to HERA data.
We begin the analysis by showing the structure functions $F_L$ and $F_2$ computed with the finite energy correction, Eq.~\eqref{eq:Final_cs}, divided by the corresponding results obtained without the finite energy correction, i.e. the optical theorem result~\eqref{eq:OT_cs}. We calculate this ratio as a function of the upper limit of the quark-antiquark pair invariant mass $M_{\mathrm{max}}^2$ for different $Q^2$, to demonstrate the effect of the invariant mass cutoff.  The results are shown in Fig.\,\ref{fig:Light1Dplot}. The finite energy correction is found to have a larger effect in $F_2$ than in $F_L$. This is because $F_2$ is dominated by the contribution from the transverse photon, in which case the aligned jet limit is numerically important.  In this limit one of the quarks carries a large fraction $z\sim 1$ (or $1-z\sim 1$) of the $\gamma^*$ plus momentum. Consequently the final state invariant mass is typically large, and having a finite upper limit $M_\mathrm{max}^2$ has a larger effect for the transverse polarization.

In the limit $M_\mathrm{max}^2\to \infty$, the cross section matches the optical theorem result as expected. The effect of having a finite $M_\mathrm{max}^2$ becomes more pronounced at higher $Q^2$. 
This is because the relative  transverse momenta of the partons are typically of the order $Q^2$, so that at higher $Q^2$ the invariant masses would typically be higher and cutting them off by $M_\mathrm{max}^2$ has a  larger effect. 

However, to quantify the effect on  physical cross sections, we cannot take  $M_\mathrm{max}^2$ to be a free parameter, but should take it to be given by $M_\mathrm{max}^2=W^2=Q^2/x$, which in turn depends on $Q^2$ and $x$. Thus we cannot directly read the phenomenological impact from the previous plot that had a fixed $Q^2$. 
Instead, in Fig.~\ref{fig:LightHeatmap} we  quantify the effect due to the finite energy constraint by the ratio to the optical theorem result, defined as  $1 - F_{L,2}^{FE}/F_{L,2}^{OT}$, as a function of $x$ and $Q^2$. The effect of the finite energy correction on the cross section is found to be largest in the region where $Q^2$ is small  ($Q^2 \sim 1$ GeV$^2$) and $x$ is relatively large ($x \sim 0.01$). This is expected, as it corresponds to a lower photon-nucleon center-of-mass energy $W$. The effect is significantly larger for $F_2$ than for $F_L$, reaching  up to 7\% compared to 0.9\%. In particular, the effect on $F_2$ is comparable to, and even larger than, experimental uncertainties in the final combined HERA structure function data~\cite{H1:2015ubc}. This can also be compared to the next-to-eikonal corrections in dijet production in DIS, which are estimated to be  of similar magnitude, around $\sim \mathcal{O}(10\%)$~\cite{Agostini:2024xqs}. Numerical values for the relative suppression due to the finite energy correction are available as a Supplementary Material to this paper.

\begin{figure*}[tb!]   % or [htbp]
    \centering
    \subfloat{\includegraphics[width=0.5\textwidth]{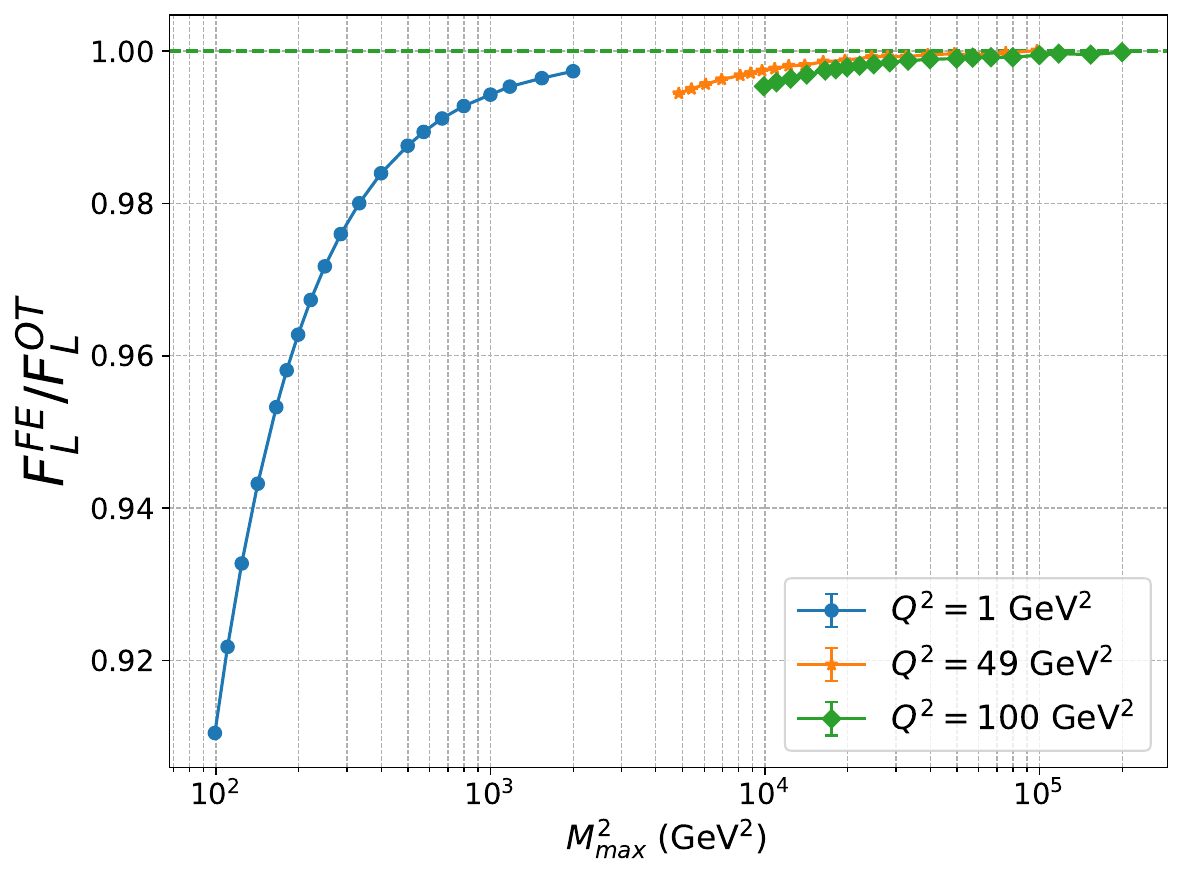}}\hfill
    \subfloat{\includegraphics[width=0.5\textwidth]{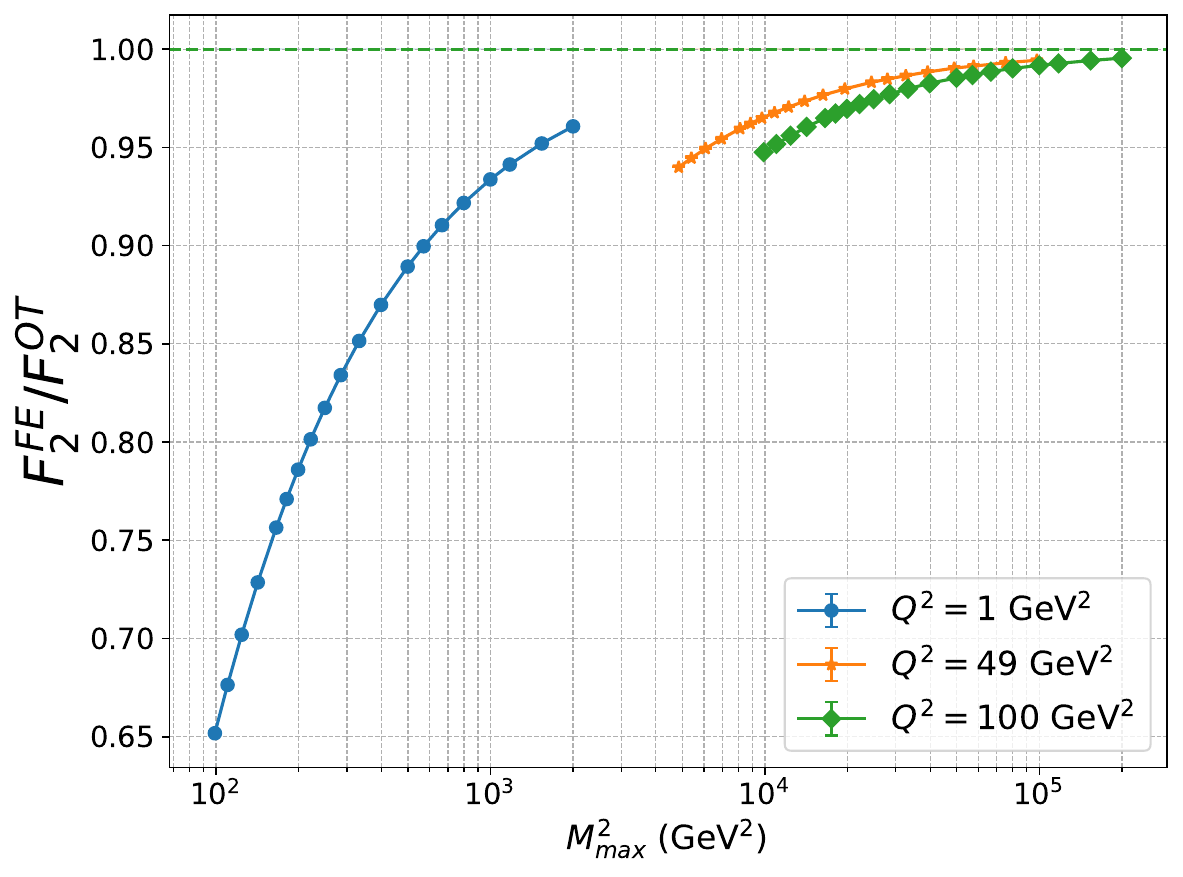}}
    \caption{Phenomenological impact of the finite energy correction on $F_L$ (left) and $F_2$ (right) for charm quarks as ratios for selected values of $Q^2$.}
    \label{fig:Heavy1Dplot}
\end{figure*}
\begin{figure*}[tb!]   % or [htbp]
    \centering
    \subfloat[]{\includegraphics[width=0.5\textwidth]{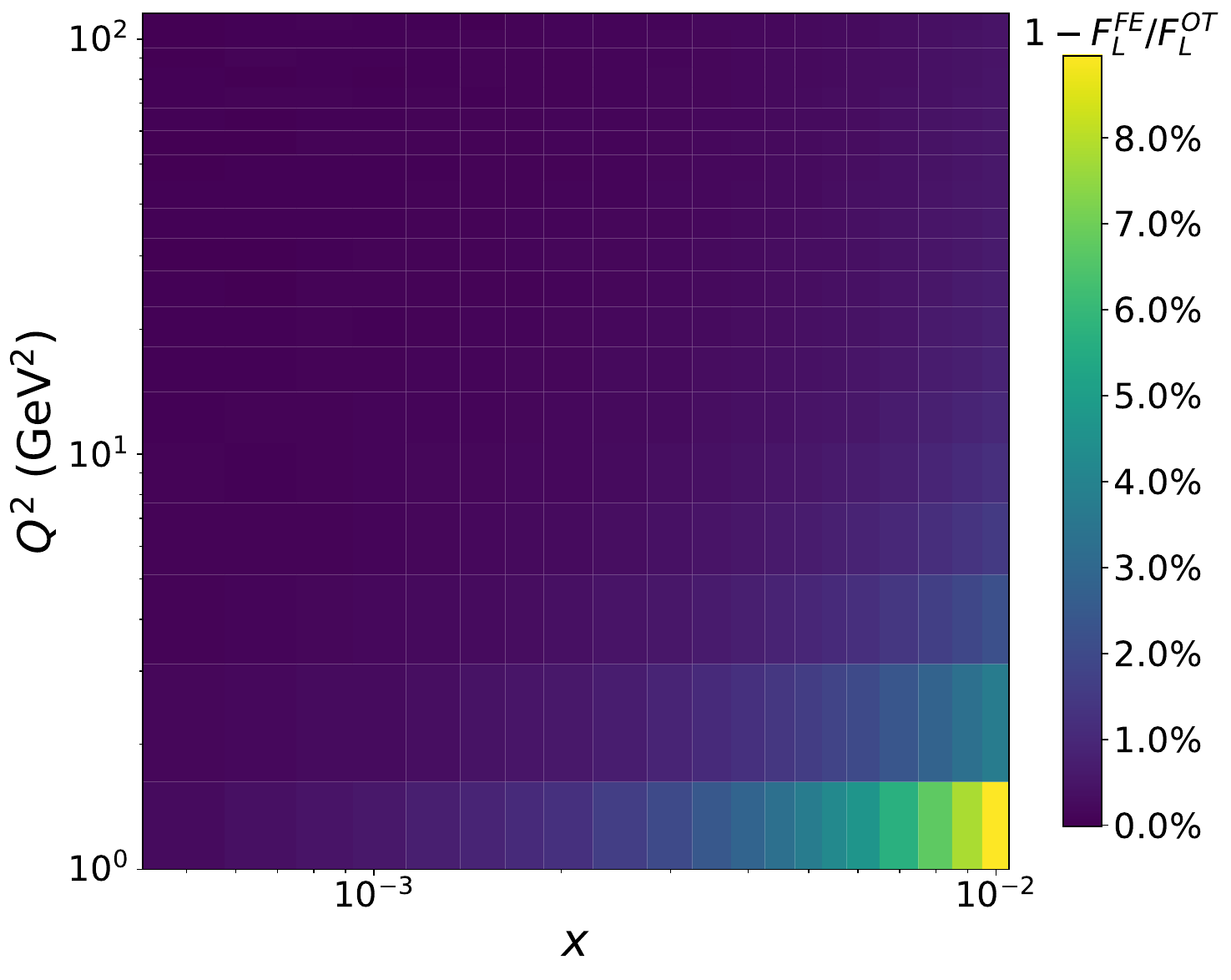}}\hfill
    \subfloat[]{\includegraphics[width=0.5\textwidth]{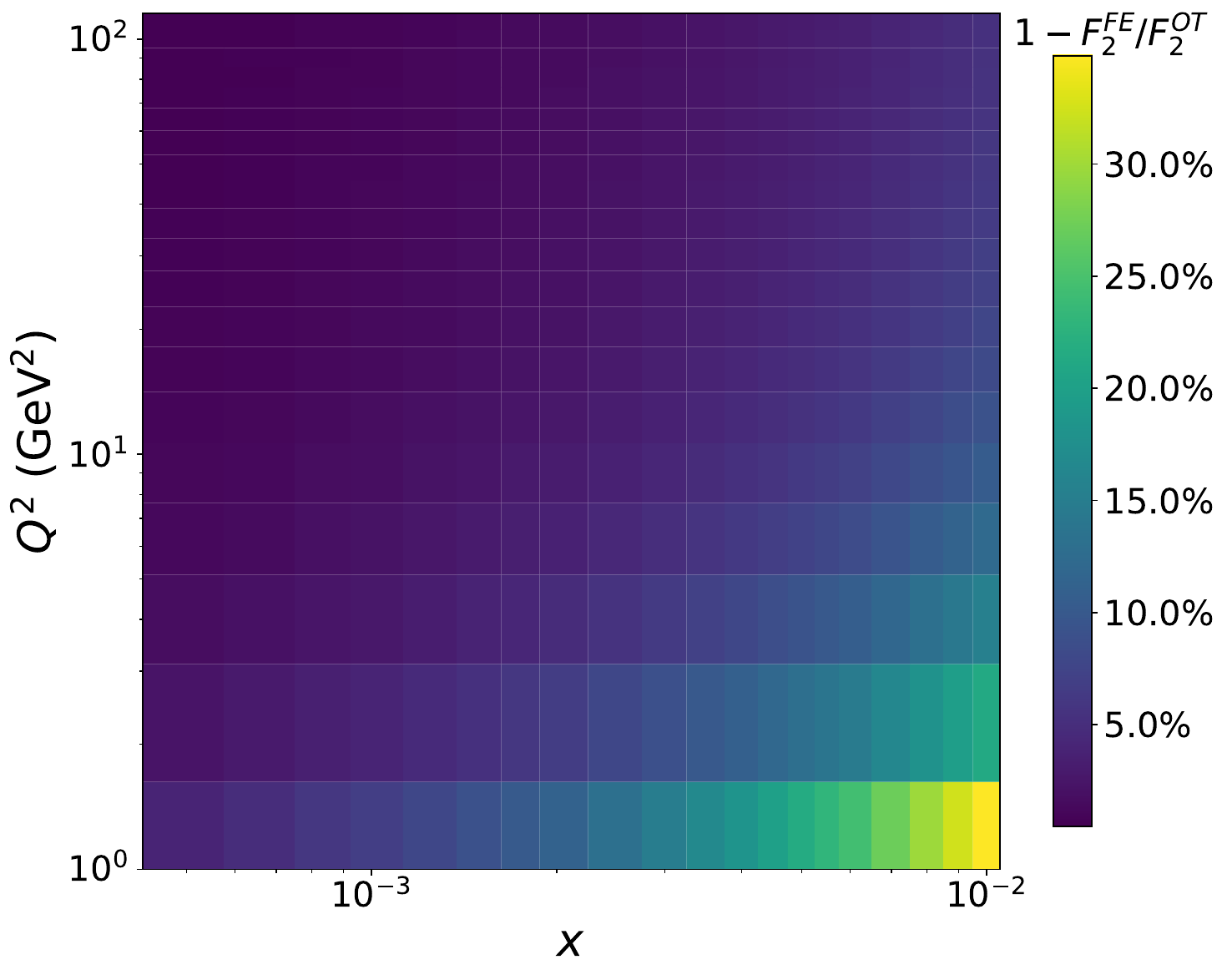}}
    \caption{Phenomenological impact of the finite energy correction on $F_L$ (left) and $F_2$ (right) in the $x$-$Q^2$ plane for charm quark production.}
    \label{fig:HeavyHeatmap}
\end{figure*}

\subsection{Heavy quarks}
For heavy quarks, we use  $m_f = 1.4$ GeV, corresponding  to the charm quark mass. We again first present the ratio of the structure functions computed with and without the finite energy  correction, as a function of the maximum allowed final-state invariant mass $M_\mathrm{max}^2$. The results for $F_{L,c}$ and $F_{2,c}$ are shown in Fig.~\ref{fig:Heavy1Dplot}. 
The effect is significantly larger than for light quark production shown in Fig.~\ref{fig:Light1Dplot}, although the overall systematics remains similar: the correction impacts $F_{2,c}$ more than  $F_{L,c}$, and for a fixed $M_\mathrm{max}^2$ is larger at higher $Q^2$.

As for light quarks, we  quantify the correction in the DIS kinematical $x,Q^2$ plane by computing the ratio between the finite energy corrected and the optical theorem results. 
Recall that  now we again take $M_\mathrm{max}^2=W^2$ rather than keeping it as a fixed number, which means that the invariant mass cutoff will depend on both $x$ and $Q^2$ via $W^2$. The results for  $F_{L,c}$ and $F_{2,c}$ are shown in Fig.~\ref{fig:HeavyHeatmap}. 
As noted above, the  correction has a substantial effect especially in the relatively large $x$ and small $Q^2$ region. This suppression is much stronger for heavy quark production  than for light quarks, because the final state invariant mass includes an additional term $m_f^2/[z(1-z)]$, see Eq.~\eqref{eq:invariant_mass_TRF}. Consequently, the invariant mass in heavy quark production is always larger than for light quarks at the same $Q^2$, making the finite upper bound on $M_{q\bar q}^2$  numerically more significant. 

Focusing on the relatively large $x\sim 0.01$ and small $Q^2\sim 1\,\mathrm{GeV}^2$ region,  the finite energy correction reduces $F_{L,c}$ by about $8\%$,  roughly 10 times more than for light quarks. For $F_{2,c}$, the suppression is approximately $35\%$, about 5 times larger than for light quarks. 
The weaker mass dependence observed for $F_2$ is due to the fact that the heavy quark mass eliminates large-invariant-mass aligned-jet configurations which are  strongly suppressed by the finite energy constraint. Since these configurations do not contribute to $F_L$, the effect of the constraint on $F_L$ is overall smaller, but more strongly dependent on the quark mass.

At all $x,Q^2$ considered here (and included in fits to extract the initial condition for the BK evolution~\cite{Albacete:2009fh, Albacete:2010sy, Casuga:2023dcf, Gao:2025dkn, Mantysaari:2018zdd, Beuf:2020dxl,Hanninen:2025iuv, Hanninen:2022gje,Casuga:2025etc}),
the finite energy correction has a numerically substantial impact  on charm production, and thus is expected to influence studies of $F_{2,c}$ and $F_{L,c}$. For example, in Ref.~\cite{Albacete:2010sy} it was found to be impossible to simultaneously describe both the total DIS cross section and the charm production data in a leading order calculation without  a separate normalization factor to suppress  charm quark production  (see however Refs.~\cite{Hanninen:2022gje,Casuga:2025etc} for recent NLO developments). This raises the prospect that the inclusion of the finite energy constraint would make even simultaneous fits to heavy and light quark cross sections work more naturally already at leading order.

\section{Conclusions}\label{sec:Conclusion}
The availability of precise HERA data for the proton structure function $F_2$, and the expected precision of the upcoming EIC, creates a need for precise calculations of proton and nuclear structure functions in the dipole picture. Extensive efforts have already been made to push calculations beyond leading order and in exploring sub-eikonal corrections. In this work, we incorporate exact kinematics by imposing a finite energy constraint $M_{q\Bar{q}}^2 \leq W^2$ on the leading order dipole picture cross section. Equation~\eqref{eq:Final_cs} provides the inclusive DIS cross section with the finite energy correction, expressed in a way that requires both quadrupole and dipole correlators. This procedure shows that finite energy, moderate $x$, effects can be taken into account even if one maintains the picture of eikonal interaction with the target, parametrized in terms of the conventional Wilson line correlators. It would be interesting to study in more detail the relation of this approach to the generalization of the dipole picture at moderate $x$ proposed in Ref.~\cite{Boussarie:2020fpb}.

We estimated the numerical effect of the finite energy correction in HERA kinematics  using the $\text{MV}$ model for the dipole and evaluating the quadrupole via the Gaussian approximation. For light quarks, the finite energy correction reduced the structure function $F_L$ up to $0.9\%$ and $F_2$ up to $7\%$ compared to calculations based on the optical theorem at $x\sim 0.01$ and $Q^2\sim 1\,\mathrm{GeV}^2$. For heavy quarks ($m_f = 1.4$ GeV), the impact was much larger, up to $9\%$ for $F_{L,c}$ and $35\%$ for $F_{2,c}$ in the same kinematics. 
This pronounced effect for heavy quarks could potentially explain the tension between the total cross section and charm quark production data when fitting the initial condition for the BK evolution to HERA data~\cite{Albacete:2010sy}. In future work, we plan to incorporate the finite energy constraint into the Bayesian inference framework of Ref.~\cite{Casuga:2023dcf} used to determine the BK initial condition, and to derive the corresponding constraint for the dipole picture DIS cross section at next-to-leading order accuracy.

\begin{acknowledgements}
The authors would like to thank Felix Hekhorn for helpful discussions regarding the development of the code used in the project. This work was supported by the Research Council of Finland, the Centre of Excellence in Quark Matter (project 346324 and 364191), and projects 338263 and 359902, and by the European Research Council (ERC, grant agreements No. ERC-2023-101123801 GlueSatLight and No. ERC-2018-ADG-835105 YoctoLHC). M.B would like to thank the \textit{Magnus Ehrnrooth foundation} for supporting this work. The content of this article does not reflect the official opinion of the European Union and responsibility for the information and views expressed therein lies entirely with the authors. 
Computing resources from CSC – IT Center for Science in Espoo, Finland and from the Finnish Computing Competence Infrastructure  (persistent identifier \texttt{urn:nbn:fi:research-infras-2016072533}) were used in this work.  
\end{acknowledgements}
\appendix
\section{Gaussian approximation}\label{sec:Appendix}
For an arbitrary set of transverse coordinates $\mathbf{x},~\mathbf{y},~\mathbf{u},~\mathbf{v}$, the quadrupole in the Gaussian approximation is given by~\cite{Dominguez:2011wm} 
\begin{equation}
    \begin{split}
        S^{(4)}(\mathbf{x},\mathbf{y},&\mathbf{u},\mathbf{v}) = e^{\frac{-C_F}{2}\left[\Gamma(\mathbf{x}, \mathbf{y}) - \Gamma(\mathbf{u},\mathbf{v})\right]}\\ 
        &\cross\Bigg[\left(\frac{\sqrt{\Delta} + F_1}{2\sqrt{\Delta}} - \frac{F_2}{\sqrt{\Delta}}\right)e^{\frac{\nc}{4}\mu^2 \sqrt{\Delta}} \\
        &+ \left(\frac{\sqrt{\Delta} - F_1}{2\sqrt{\Delta}} + \frac{F_2}{\sqrt{\Delta}}\right)e^{\frac{-\nc}{4}\mu^2 \sqrt{\Delta}}  \Bigg]\\
        &\cross e^{\mu^2\left(\frac{-\nc}{4}F_1 + \frac{1}{2\nc}F_2\right)}
    \end{split}
\end{equation}
where 
\begin{align}
\Gamma(\mathbf{x},\mathbf{y})
  &= -\frac{2}{C_F}\,f(|\mathbf{x}-\mathbf{y}|), \label{eq:Gamma} \\[8pt]
F(\mathbf{x},\mathbf{y},\mathbf{u},\mathbf{v})
  &= \frac{1}{\mu^2 C_F}\Big(
       f(|\mathbf{x}-\mathbf{u}|) 
     + f(|\mathbf{y}-\mathbf{v}|) \nonumber \\[-3pt]
  &\qquad\qquad
     - f(|\mathbf{x}-\mathbf{v}|) 
     - f(|\mathbf{y}-\mathbf{u}|)
     \Big), \\[8pt]
\Delta(\mathbf{x},\mathbf{y},\mathbf{u},\mathbf{v})
  &= F^2(\mathbf{x},\mathbf{u},\mathbf{y},\mathbf{v}) \nonumber \\[-3pt]
  &\quad
     + \frac{4}{\nc^2}\,
       F(\mathbf{x},\mathbf{y},\mathbf{u},\mathbf{v})\,
       F(\mathbf{x},\mathbf{v},\mathbf{y},\mathbf{u}).
\end{align}
and we define
\begin{equation}
    \begin{split}
        F_1 &:= F\left(\mathbf{x},\mathbf{u}, \mathbf{y},\mathbf{v}\right),\\
        F_2 &:= F\left(\mathbf{x},\mathbf{y},\mathbf{u},\mathbf{v} \right), \\
        F_3 &:= F\left(\mathbf{x}, \mathbf{v}, \mathbf{y}, \mathbf{u}\right).
    \end{split}
\end{equation}

The function $\Gamma(\xt,\yt)$ in Eq.~\eqref{eq:Gamma} is related to the two-point function as $S^{(2)}(\xt,\yt)=\exp\left\{-\frac{\cf}{2} \Gamma(\xt,\yt)\right\}$, and using the MV model dipole~\eqref{eq:dipole} we obtain
\begin{equation}\label{eq:MV_model}
    \begin{split}
        f(\abs{\mathbf{x}-\mathbf{y}}) &= -\frac{(\abs{\mathbf{x}-\mathbf{y}}^2 Q_{s0}^2)^\gamma}{4} \\
        & \times \log\left(\frac{1}{\abs{\mathbf{x}-\mathbf{y}}\Lambda_{QCD}} + e\cdot e_c\right).
    \end{split}
\end{equation}

\bibliography{refs}
\bibliographystyle{JHEP-2modlong}

\end{document}